# Single artificial-atom lasing

O. Astafiev[1,2], K. Inomata[2], A. O. Niskanen[3,4], T. Yamamoto[1,2,3], Yu. A. Pashkin[1,2], Y. Nakamura[1,2,3] & J. S. Tsai[1,2,3]

[1]*NEC Nano Electronics Research Laboratories. Tsukuba, Ibaraki, 305-8501, Japan*

[2]*The Institute of Physical and Chemical Research (RIKEN), Wako, Saitama 351-0198, Japan*

[3]*CREST-JST, Kawaguchi, Saitama 332-0012, Japan*

[4]*VTT Technical Research Center of Finland, Sensors, POB 1000, 02044 VTT, Finland*

**Solid-state superconducting circuits[1-3] are versatile systems in which quantum states can be engineered and controlled. Recent progress in this area has opened up exciting possibilities for exploring fundamental physics as well as applications in quantum information technology; in a series of experiments[4-8] it was shown that such circuits can be exploited to generate quantum optical phenomena, by designing superconducting elements as artificial atoms that are coupled coherently to the photon field of a resonator. Here we demonstrate a lasing effect with a single artificial atom—a Josephson-junction charge qubit[9]—embedded in a superconducting resonator. We make use of one of the properties of solid-state artificial atoms, namely that they are strongly and controllably coupled to the resonator modes. The device is essentially different from existing lasers and masers; one and the same artificial atom excited by current injection produces many photons.**

Conventional lasers and masers consist of many atoms that are weakly coupled to a cavity owing to the tiny size of natural atoms. Nevertheless, by using a tightly confined cavity mode, coherent interaction of a single atom and the cavity can be achieved: the atom-cavity interaction time becomes shorter than the photon lifetime or the atom coherence time[10, 11]. Such a strong coupling regime results in a qualitatively new feature: the vanishing of the pumping threshold that has been experimentally realized in single-atom masers and lasers [12, 13]. On the other hand, quantum systems with artificial atoms allow one to easily make the interaction time much shorter than the coherence time, as it has been demonstrated recently[4-6,14,15]. Furthermore, controllable interaction with a single cavity mode together with a fast mechanism of population inversion gives a possibility to realize a lasing regime with many photons generated by one and the same atom[13, 16, 17].

In this work we demonstrate lasing action in a maser operation based on a single Josephson-junction charge qubit with a population inversion mechanism provided by single-electron tunnelling events[18]. Alternative lasing schemes with superconducting qubits have been discussed elsewhere. 19, 20.

The artificial-atom maser consists of a resonator and a charge qubit coupled to it (Fig. 1). We fabricate a transmission-type half-wavelength coplanar-waveguide resonator using a 200-nm thick Nb film (Fig. 1**c**). It has a bare resonance frequency $\omega_0/2\pi = 9.899$ GHz and a quality factor $Q = 7.6 \times 10^3$. The corresponding photon decay rate is $\kappa/2\pi = 1.3$ MHz. The qubit is fabricated by three-angle shadow evaporation of Al close to the end of the resonator, where the electric field is nearly maximal. The qubit[9] is well described by two charge states, $|0\rangle$ and $|2\rangle$, differing by one Cooper pair (consisting of two electrons) in the island, and is characterized by the Josephson energy $E_J$ and the single-electron charging energy $E_C$. The electrostatic energy difference $\varepsilon = 4 E_C (n_g - 1)$ between the two states is controlled by the normalized gate charge $n_g = C_g V_g/e$, where $V_g$ is the gate voltage, $C_g$ the gate capacitance and $e$ the electron charge. The qubit eigenenergy follows $\Delta E = \sqrt{\varepsilon^2 + E_J^2}$ (top right panel of Fig. 1**d**). As the qubit is coupled to the resonator through the electric field ($\propto a^+ + a$, where $a^+$ and $a$ are the photon creation and annihilation operators, respectively), the hamiltonian of the qubit-resonator system reads:

$$H = -\frac{1}{2}(\varepsilon\sigma_z + E_J\sigma_x) + \hbar\omega_0\left(a^+a + \frac{1}{2}\right) + \hbar g_0(a^+ + a)\sigma_z. \qquad (1)$$

The first term represents the qubit; $\sigma_z$ and $\sigma_x$ are the Pauli matrices. The second term describes the resonator. The interaction between the qubit and the resonator gives the third term, and is characterized by the coupling strength $g_0$. The value of $g_0/2\pi$ is found to be 80 MHz from fitting the dispersion curve observed in the transmission through the resonator when the qubit is biased at $\varepsilon = 0$ (Fig. 2**a**).

To create population inversion in the qubit we introduce a drain electrode connected to the island via a tunnel junction with the resistance $R_b$ of 1.0 MΩ (Figs. 1**a**, **b**). The drain electrode is voltage biased at a voltage $V_b$ above $(2\Delta + E_C)/e$, which is required to extract two electrons from the island by breaking a Cooper pair (where $\Delta$ is the superconducting gap energy; $\Delta/h \approx 55$ GHz) [21, 22]. As a result, $|2\rangle$ decays into $|0\rangle$ via two sequential single-electron tunnelling events in the incoherent process $|2\rangle \rightarrow |1\rangle \rightarrow |0\rangle$ with rates $\Gamma_{21}$, $\Gamma_{10} \approx (eV_b \pm E_C)/e^2 R_b$ (positive sign for the former), respectively (bottom left panel of Fig. 1**d**). Therefore, the 'atom' is pumped into $|0\rangle$ state regardless of the sign of $\varepsilon$. At $\varepsilon = 0$, a Cooper pair tunnels across the Josephson junction from the ground to the island ($|0\rangle \rightarrow |2\rangle$) without changing its energy. Thus, the so-called Josephson-quasiparticle (JQP) cycle involving the three charge states continues and results in a pronounced current peak[21,22]. For $\varepsilon \gg E_J$, the upper eigenstate of the qubit is nearly the $|0\rangle$ state, and the single-electron tunnelling process creates population inversion with an effective rate $\gamma = \Gamma_{21}\Gamma_{10}/(\Gamma_{21}+\Gamma_{10})$. For $V_b = 0.65$ mV used in the measurement below, $\gamma \approx 2.0\times10^9$ s$^{-1}$ ($\gamma/2\pi \approx 320$ MHz) which is much larger than $\kappa$.

When $\Delta E$ is adjusted to $\hbar\omega_0$ the energy quantum of the qubit is transferred into the resonator as a photon, accompanied by a Cooper pair tunnelling across the Josephson junction ($|0, N\rangle \rightarrow |2, N+1\rangle$; $|n, N\rangle$ represents a state with $n$ excess electrons in the qubit island and $N$ photons in the resonator). Completed by the pumping mechanism $|2, N+1\rangle \rightarrow |1, N+1\rangle \rightarrow |0, N+1\rangle$, the photon-assisted JQP cycle proceeds repeatedly with increasing $N$, and $N$ reaches the balance between the photon generation and the loss of the resonator. The coupling between $|0, N\rangle$ and $|2, N+1\rangle$ states is enhanced by a factor of $\sqrt{N+1}$; the photon field stimulates the photon generation process, which is analogous to stimulated emission in conventional lasers. However, in our case the photons are generated by one and the same atom. In conventional lasers, the ratio $\beta$ of spontaneous decay rate into the lasing mode to the total spontaneous decay rate is very low. Therefore, high pumping rate (above lasing threshold) is required to achieve lasing. However in our system, a single atom efficiently coupled to a single-mode cavity with $\beta$ close to unity, the threshold no longer exists and lasing takes place at any weak pumping rate[13, 16].

In Fig. 2**b**, emission power spectral density from the resonator (upper panel) is shown together with the current through the qubit (lower panel) as a function of $\varepsilon$. The observed current peak at $\varepsilon = 0$ is due to the JQP process. On the right slope of the JQP peak ($\varepsilon > 0$; the emission side), two small current peaks ($I_p \approx 0.1$ nA above JQP peak) appear. Correspondingly, we observe strong emission shown as two "hot spots" in the upper panel. The position of the first current peak and the hot spot corresponds to $\varepsilon/2\pi \sim 7\pm2$ GHz. Although the hot spot is rather broad, it is located consistently with the condition $\Delta E = \hbar\omega_0$ ($\varepsilon = 8.3$ GHz). (Although the emission takes place in a wide range of the magnetic flux $\Phi$, the data shown here is obtained at $\Phi = 0.38\ \Phi_0$ ($\Phi_0$ is the flux quantum), where $E_J/h \approx 5.4$ GHz). Because of finite $\varepsilon$, the effective coupling strength at the resonance is reduced to $g/2\pi = (g_0/2\pi)(E_J/\hbar\omega_0) \approx 44$ MHz.) One possible interpretation of the presence of the second hot spot is the two-photon resonance[18] expected at $\varepsilon = 19$ GHz. Note that the emission takes place only when the drain electrode is biased in the range 0.57 mV $\leq V_b \leq$ 0.71 mV, where the JQP cycle is the dominant current carrying process. Note also that on the absorption side, $\varepsilon < 0$, the microwave de-amplification is expected at $\Delta E = \hbar \omega_0$ and is indeed observed, though it is not shown here.

Figure 2**c** shows the emission spectrum at one of the hot spots. The frequency of the intense emission is shifted by $\sim -0.7$ MHz from the resonator frequency. The emission peak is unstable, showing low-frequency fluctuations, which can be attributed to the low frequency charge noise. However, it is roughly confined within the envelope drawn by the black curve. The total emission power within the envelope is estimated to be $W = 7 \times 10^{-16}$ W, which corresponds to $N = 2\,(W/\hbar\omega_0)/\kappa \approx 30$ photons in the resonator. (The factor 2 comes from equal probability for the photons to escape from each end of the resonator: the number 30 may be underestimated, as the resonator internal loss is not accounted for.) The large number of photons accumulated in the single-mode resonator indicates lasing effect, together with the linewidth narrower than $\kappa$ as well as $\Gamma_{21}$. However, the linewidth is still much wider than the quantum limit given by the Schawlow-Townes formula[23] $\kappa/(2N)$ (of order of $2\pi \times 10$ kHz), which means that it is broadened by some other mechanism, e.g., charge fluctuations. The $\beta$-factor is estimated as a ratio of photon escape rate over the photon assisted Cooper-pair tunnelling rate $\beta > (N\kappa)/(I_p/2e) = 0.4$. It supports our picture of high lasing efficiency.

To additionally prove the lasing action of our device, we study the amplification of an external microwave. Figure 3**a** shows the normalized power and phase of coherent radiation output from the resonator. The blue curves show an ordinary transmission through the resonator when the qubit is biased away from the hot spots, and, as expected, the amplitude exhibits a lorentzian shape. The red curve demonstrates amplification of the drive microwave: at the hot spots, the transmission peak is enhanced on the low-frequency side of the bare resonant peak and slightly shifted towards lower frequencies in respect to the emission peak. At frequency $\delta\omega_{drive}/2\pi \sim -0.6$ MHz ($\delta\omega_{drive} \equiv \omega_{drive} - \omega_0$), the amplification switches to attenuation accompanied by the phase drop. One possible interpretation of the frequency shift and the phase drop in Figs 2**c** and 3**a** is that they are signatures of qubit-resonator coupling $g(N+1)^{1/2}$ that makes the system non-linear in photon fields $N^{1/2}$. The resonance frequency and consequently the amplification and emission peaks are expected to split by approximately $\pm g/2N^{1/2}$. The observed amplification peak is shifted by $\sim -1$ MHz, which is of the order of the expected value. However, the peaks shifted to positive frequencies are not observed. The phase shift accompanying an amplification peak of a narrow band amplifier should also be additionally affected by the nonlinear term $\sim N^{1/2}$ and therefore drops on the right slope of the amplification peak, where $N$ is suppressed.

Next we study emission spectrum under the external driving microwave, expecting "injection locking" effects[22]. The red curve in Fig. 3**b** exemplifies the emission power spectrum at the hot spot when the external drive power $P_{drive}$ corresponding to six photons in the resonator ($N^* \equiv (P_{drive}/\hbar\omega_0)/\kappa \approx 6$) is applied. The driven emission (red curve) reproduces the shape of the drive signal (blue curve) at frequency $\omega_{drive}/2\pi$ ($\delta\omega_{drive}/2\pi = -0.5$ MHz), while the emission is suppressed at $\delta\omega \neq \delta\omega_{drive}$. This is consistent with the expected locking mechanism of the emission. The red peak is also much higher than the blue one which is the transmitted spectrum measured with the same drive power and at $\varepsilon/h = 40$ GHz (outside the hot spots). We measured the injection locking in the range of $\delta\omega_{drive}/2\pi$ from $-1.5$ MHz to $0.5$ MHz and found that locking takes place at higher power for larger detuning from the emission peak maximum. The spectrum strongly depends on $P_{drive}$ (Fig. 3**c**). When $N^*$ exceeds 1, the emission line shrinks to the drive frequency with the width limited by the measurement bandwidth (100 kHz) and amplitude fluctuations in the locked signal are suppressed. The injection locking effect resulting in frequency stabilization and emission narrowing additionally proves the lasing action.

We have demonstrated a lasing effect in the simplest possible geometry – one 'atom' coupled to a resonator. The physical simplicity and controllability makes it especially attractive for studying fundamental laser properties. We expect that the artificial-atom masers can be used as on-chip microwave sources and microwave amplifiers.

**Acknowledgments**   We are grateful to A. Zagoskin, A. Smirnov, L. Murokh, S. Kouno, A. Tomita, and A. Clerk for useful discussions.

**Competing Interests** The authors declare that they have no competing financial interests.

**Correspondence** Correspondence and requests for materials should be addressed to O. Astafiev (astf@frl.cl.nec.co.jp).

**Figure captions**

**Figure 1 Single artificial-atom maser and lasing mechanism**. **a**, Schematic representation of the circuit. **b**, Scanning-electron micrograph of the qubit. The Josephson charge qubit consists of a superconducting Al island, with the charging energy $E_C = e^2/2C_\Sigma = h \times 20$ GHz ($C_\Sigma$ is the total capacitance of the island), connected to the ground through two Josephson junctions with a SQUID geometry so that the effective Josephson energy $E_J$ is controlled by a magnetic flux $\Phi$ through the loop. A voltage-biased drain electrode is connected to the island via a 1.0-M$\Omega$ tunnel junction. A tail of an Al strip (see also **c**) forms another tunnel junction with estimated capacitance of $C_r \sim 200$ aF, defining the qubit-resonator coupling. The small junction conductance $\sim(200$ k$\Omega)^{-1}$ is irrelevant for the unbiased junction. **c**, Micrograph of the left end of the Nb coplanar waveguide resonator. At each end of the resonator, the central line is capacitively coupled to the external microwave line with a characteristic impedance of 50 $\Omega$. The qubit is fabricated close to the end of the $\sim$ 6.24-mm long resonator. Bright stripes on top of the SiO$_2$ insulating layer are the qubit dc bias lines. An Al strip extends from the resonator towards the qubit for realizing strong capacitive coupling. **d**, Energy band diagram of the qubit (top right) and the lasing mechanism (bottom left). For $\varepsilon > 0$, population inversion is created by two sequential single-electron tunnelling events ($|2\rangle \rightarrow |1\rangle \rightarrow |0\rangle$) from the island to the drain.

**Figure 2 Emission from the self-running maser. a**, Transmission spectrum through the resonator measured with a weak microwave power ($P = -138$ dBm) as a function of the magnetic flux $\Phi$ in the SQUID loop. The average number of photons in the resonator is kept below 0.3. The detuning $\delta\omega/2\pi \equiv (\delta\omega - \delta\omega_0)/2\pi$ is the difference between the probe frequency $\omega/2\pi$ and the resonator frequency

$\omega_0/2\pi$ (= 9.899 GHz). The qubit is biased at $\varepsilon = 0$ and $V_b = 0$, so that the qubit energy $\Delta E$ equals to $E_J$ and there is no current injection. The observed dispersion curve is reproduced by $\delta\omega = \left[(E_J - \hbar\omega_0) \pm \sqrt{(E_J - \hbar\omega_0)^2 + 4g_0^2}\right]/2\hbar$ with $E_J = E_{J0}$ cos$|\pi\Phi/\Phi_0|$ (where $\Phi_0$ is the flux quantum), $E_{J0}/h = 13.7$ GHz and $g_0/h = 80$ MHz (red dashed lines). **b**, Emission power spectrum $S$ from the resonator (upper panel) together with the current $I$ through the qubit (lower panel) as a function of $\varepsilon$ or $n_g$. Population inversion mechanism due to the JQP process is now activated with $V_b = 0.65$ mV. The Josephson energy of the qubit is reduced to $E_J/h = 5.4$ GHz by applying a magnetic flux $\Phi = 0.38\ \Phi_0$. The emission is seen as two "hot spots", and the corresponding current peaks appear on the right slope of the JQP peak ($\varepsilon > 0$). This double hot spot feature is reproduced around every charge degeneracy point between $|n\rangle$ and $|n+2\rangle$, periodically in $n_g$. However, in another sample with lower $\gamma$, we observed a single hot spot with lower emission power. **c**, Emission power spectrum $S$ at one of the hot spots taken at $\varepsilon/h = 7$ GHz (red curve). The black curve is an eye-guide envelope of the emission peak. The background level originates from the amplifier which has the noise temperature of 10 K.

**Figure 3 Microwave amplification and injection locking. a**, Normalized power (upper panel) and phase (lower panel) of the coherent radiation from the resonator as a function of detuning of the driving microwave $\delta\omega_{drive}/2\pi$, off the hot spots (blue, $\varepsilon/h = 40$ GHz) and at the hot spot (red, $\varepsilon/h \approx 7$ GHz) with $V_b = 0.65$ mV. The amplitude is normalized to the input driving power of $P_{drive} = -135$ dBm corresponding to $N^* = 0.6$ at $\delta\omega_{drive} = 0$. At $\delta\omega_{drive}/2\pi \sim -0.6$ MHz the amplification regime switches to the attenuated transmission regime. The change of the regimes is also seen as a sudden phase drop. **b**, Output power spectrum $S$

under the driving microwave field at a fixed detuning $\delta\omega_{drive}/2\pi = -0.5$ MHz and with a power $P_{drive} = -125$ dBm. The blue curve is measured outside hot spots, while the red curve is taken at the hot spot ($\varepsilon/h \approx 7$ GHz). The black dashed curve is the envelope of the emission spectrum in the absence of any microwave drive (see Fig. 2**c**). **c**, Output power spectrum $S$ (colour map in log scale) as a function of the driving power for the detuning frequency $\delta\omega_{drive}/2\pi = -0.5$ MHz. The spectrum gets as narrow as the measurement bandwidth (100 kHz) when $N^* >\sim 1$.

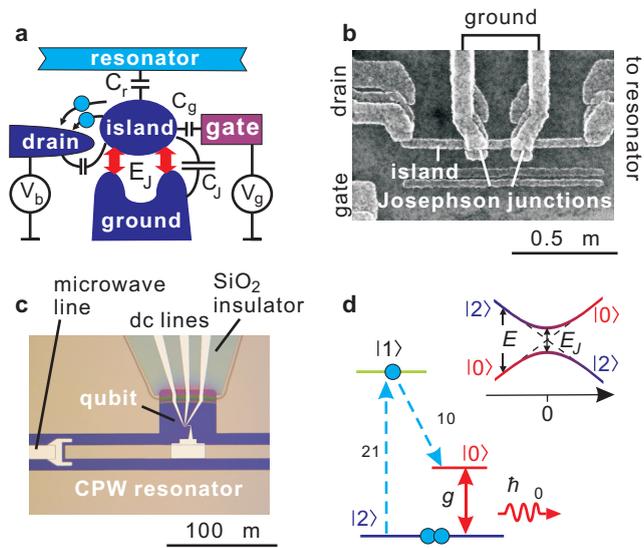

Figure 1

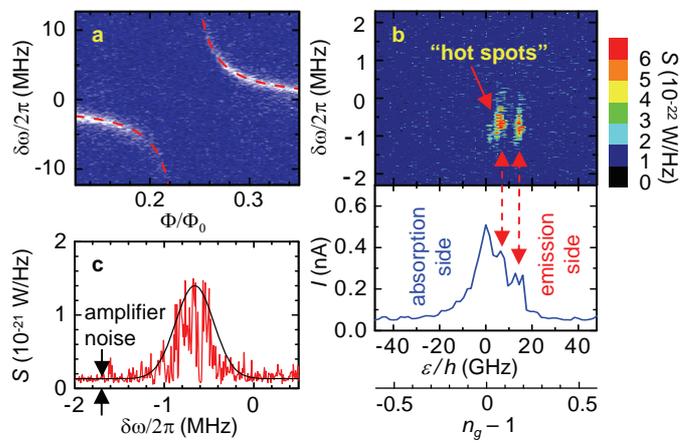

Figure 2

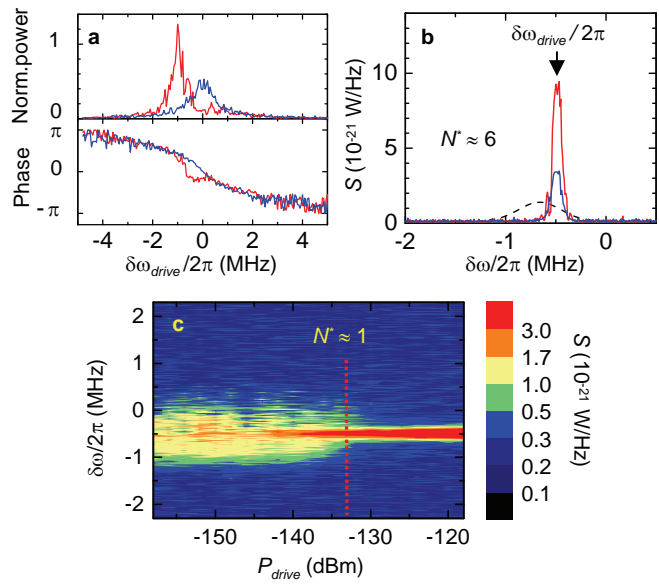

Figure 3